\documentclass{PoS}
\usepackage{wasysym}

\newcommand{\be}{\begin{equation}}
\newcommand{\ee}{\end{equation}}
\newcommand{\bea}{\begin{eqnarray}}
\newcommand{\eea}{\end{eqnarray}}

\title{Determination of Vus: Recent Input from the Lattice}

\ShortTitle{Determination of Vus: Recent Input from the Lattice}

\author{\speaker{Vittorio Lubicz}%
\\
        Dipartimento di Matematica e Fisica, Universit\`a Roma Tre and INFN, Sezione di Roma Tre\\
        E-mail: \email{lubicz@fis.uniroma3.it}}

\abstract{
The two most precise determinations of the CKM matrix element $V_{us}$ are based on the analyses of leptonic and semileptonic kaon decays. These studies also rely on the lattice QCD calculations of two hadronic parameters, namely the ratio of the kaon and pion decay constants, $f_{K^+}/f_{\pi^+} $, and the kaon semileptonic vector form factor at zero momentum transfer, $f_{+}(0)$. In this talk, I review the recent lattice results for these quantities, by showing that the sub-percent accuracy required by the phenomenological analyses has been reached by lattice QCD. As best estimates of the lattice calculations I quote $f_{K^+}/f_{\pi^+} = 1.193\ (4)$ and $f_{+}(0)=0.965\ (3)$. I also discuss some recent theoretical progress in the evaluation of the small, but phenomenologically relevant, $SU(2)$ isospin breaking corrections.
}

\FullConference{2013 Kaon Physics International Conference\\
		 Apr 29-30 and May 1st\\
		 University of Michigan, Ann Arbor, Michigan - USA}

\begin{document}
\section{Introduction}
The two most accurate sources of information on the value of the CKM matrix element $V_{us}$ are leptonic and semileptonic kaon decays. For the leptonic channel, it is convenient to consider the ratio of the kaon to the pion decay rates, so that the relevant quantities that are eventually extracted from the experimental analyses are the following two combinations of CKM matrix elements and hadronic parameters~\cite{Antonelli:2010yf}:
\be
\label{eq:exp}
\frac{|V_{us}|}{|V_{ud}|} \frac{f_{K^+}}{f_{\pi^+}} = 0.2758(5) \quad , \quad |V_{us}| \, f_{+}(0) = 0.2163(5) \ .
\ee
In Eq.~(\ref{eq:exp}), $f_{K^+}$ and $f_{\pi^+}$ are the charged kaon and pion decay constants respectively, defined in pure QCD. ~$f_{+}(0)$ is the QCD vector form factor at zero momentum transfer for the neutral kaon semileptonic decays $K^0 \to \pi^- \ell \nu$. In both cases, electromagnetic corrections to the decay rates have been already subtracted in the experimental analyses, using chiral perturbation theory supplied by approximate models to estimate the values of unknown low-energy constants (for more details, see~\cite{Antonelli:2010yf} and references therein).

The experimental accuracy displayed by the results in Eq.~(\ref{eq:exp}) is at the $2\permil$ level. As theoretical input, the determinations of Eq.~(\ref{eq:exp}) involve a precise evaluation of the short distance radiative corrections, besides the estimate of the already mentioned, long distance electromagnetic effects. This permille accuracy represents a challenge for lattice QCD calculations, which are required for a non-perturbative, model-independent determination of the ratio of pseudoscalar decay constants $f_{K^+}/f_{\pi^+}$ and of the semileptonic form factor  $f_{+}(0)$. 

In this talk, I review the status of the lattice calculations of $f_{K^+}/f_{\pi^+}$ and $f_{+}(0)$, and show that the permille accuracy, required to compete with the experimental result, is not far from being reached on the lattice. It is worth noting that, while remarkable, this result is however less surprising than it could appear at a first glance. The key observation is that both the ratio $f_{K^+}/f_{\pi^+}$ and the form factor $f_{+}(0)$ are quantities which are exactly equal to unity in the $SU(3)$ symmetric limit, so that in these cases the hadronic parameters that are evaluated on the lattice are just the deviations from the symmetric point. It then turns out that an accuracy of about $2\permil$ on $f_{K^+}/f_{\pi^+}$ and $f_{+}(0)$ corresponds to a precision of about 1.3\% and 5\% respectively on $f_{K^+}/f_{\pi^+}-1$ and $1-f_{+}(0)$. Though still challenging, this explains why such an accuracy is actually reachable by current, high-precision lattice QCD calculations.

\section{Lattice calculations of $\mathbf{f_{K}/f_{\pi}}$}
\begin{figure}
\vspace{-0.8cm}
\centering
\includegraphics[width=0.8\textwidth]{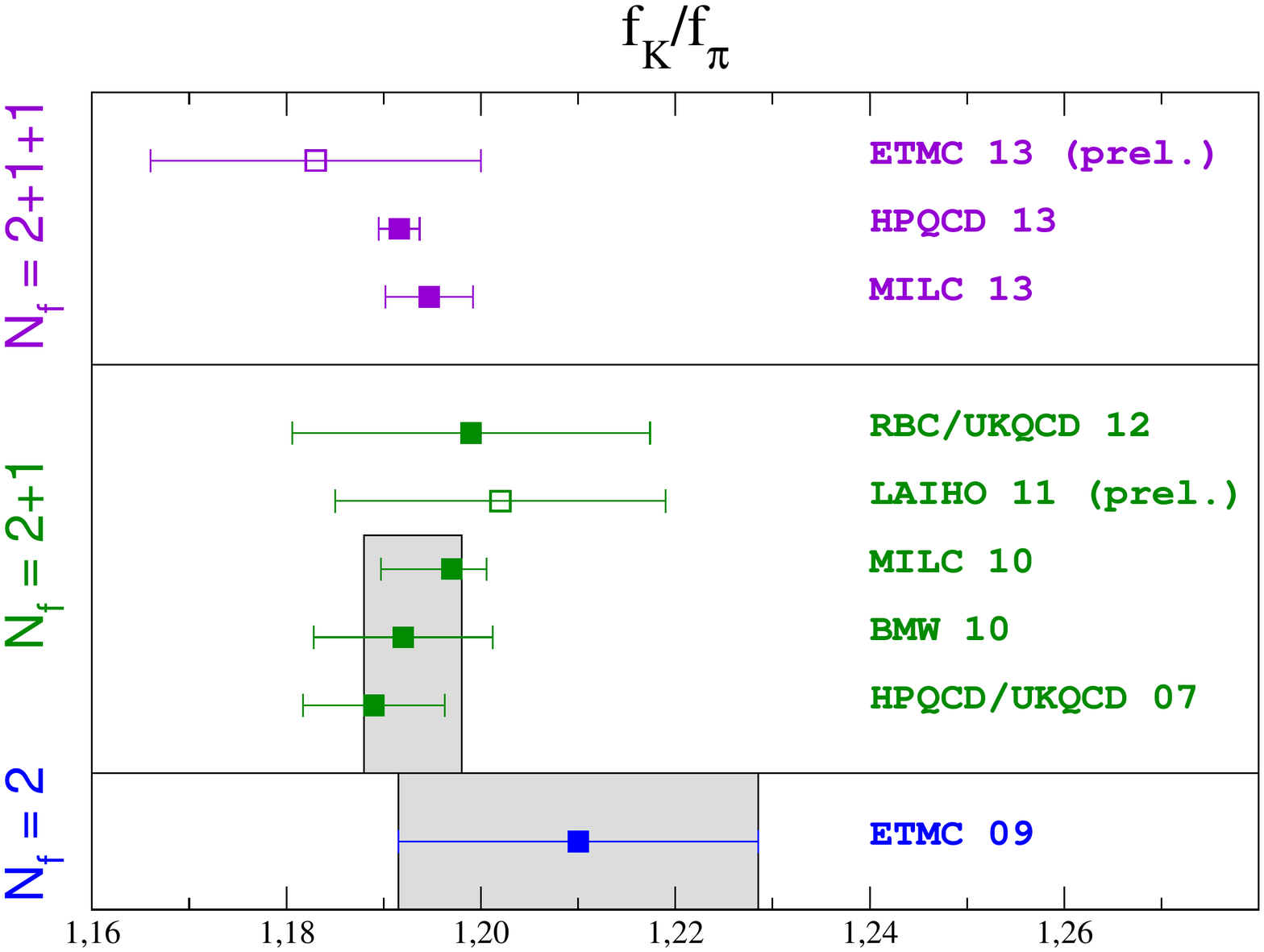}
\vspace{-1.0cm}
\caption{\it Lattice results for the ratio of decay constants $f_{K^+}/f_{\pi^+}$. The results shown in the lower side of the plot are those considered for the final averages in FLAG-1~\cite{Colangelo:2010et}. These averages are shown by the gray bands. The new results, appeared after the FLAG-1 compilation, are shown on the upper side.}
\label{fig:fkfpi}
\end{figure}

A comprehensive and critical compilation of lattice results for the ratio $f_{K}/f_{\pi}$, in the isospin symmetric limit, has been presented at the end of 2010 by the FLAG group in its first report (FLAG-1)~\cite{Colangelo:2010et}. FLAG-1 examined a large number of unquenched lattice studies, but the final averages were based on relatively few selected lattice calculations: ETMC 09~\cite{ETMC09}, HPQCD/UKQCD 07~\cite{HPQCD/UKQCD07}, BMW 10~\cite{BMW10} and MILC 10~\cite{MILC10}. These results are presented in the lower part of the plot in Fig.~\ref{fig:fkfpi}. 
The FLAG averages were quoted separately for calculations including either $N_f=2$ (degenerate up and down) or  $N_f=2+1$ (up, down, strange) dynamical sea quark effects and read:
\bea
\label{eq:fKfpi-FLAGave}
& f_{K}/f_{\pi} = 1.210\ (18) \ , & \qquad  \textrm{FLAG-1} \ \ - \ \ N_f=2 \nonumber \\
& f_{K}/f_{\pi} = 1.193\ (5) \ , & \qquad  \textrm{FLAG-1} \ \  - \ \ N_f=2 + 1 \ .
\eea
These averages are also shown  in Fig.~\ref{fig:fkfpi} by the gray bands.

After the FLAG-1 compilation, a number of new lattice results for $f_{K}/f_{\pi}$ have been presented, which are shown in the upper side of the plot in Fig.~\ref{fig:fkfpi}. The whole list of results is also collected in Table~\ref{tab:fkfpi}. 
\begin{table}[t]
\vspace{-0.5cm}
{\footnotesize
\begin{tabular*}{\textwidth}{l @{\extracolsep{\fill}} c c c c}
\hline\hline \\[-0.2ex]
  & {Ref.} & {$N_f$} & {$f_{K^+}/f_{\pi^+}$} & {$f_{K^+}$ (MeV)} 
\\[0.2ex] \hline \hline \\[-0.2ex]
ETMC 13 & \cite{ETMC13} & 2+1+1 & 1.183 (17) & 154.4 (2.1)
\\[2.0ex] \hline \\[-2.0ex]
HPQCD 13 & \cite{HPQCD13} & 2+1+1 & 1.192 (2) & 155.4 (0.3)
\\[2.0ex] \hline \\[-2.0ex]
MILC 13 & \cite{MILC13} & 2+1+1 & 1.195 (5) & 155.8 (0.6)
\\[2.0ex] \hline \hline \\[-2.0ex]
RBC/UKQCD 12 & \cite{RBC/UKQCD12} & 2+1 & 1.199 (18) $^*$ & 152 (4) $^*$
\\[2.0ex] \hline \\[-2.0ex]
LAIHO 11 (prel.) & \cite{LAIHO11} & 2+1 & 1.202 (17) & 156.8 (2.0)
\\[2.0ex] \hline \\[-2.0ex]
MILC 10 & \cite{MILC10} & 2+1 & 1.197 ($^{+4}_{-7}$) & 156.1 ($^{+0.7}_{-1.0}$)
\\[2.0ex] \hline \\[-2.0ex]
BMW 10 & \cite{BMW10} & 2+1 & 1.192 (9)$^*$ & ---
\\[2.0ex] \hline \\[-2.0ex]
HPQCD/UKQCD 07 & \cite{HPQCD/UKQCD07} & 2+1 & 1.189 (7)$^*$ & 157 (2)$^*$
\\[2.0ex] \hline \hline \\[-2.0ex]
ETMC 09 & \cite{ETMC09} & 2 & 1.210 (18)$^*$ & 158.1 (2.4)$^*$
\\[2.0ex] \hline \hline
\end{tabular*}
}
\caption{\it Lattice results for the ratio of decay constants $f_{K^+}/f_{\pi^+}$ and for $f_{K^+}$. The results from Ref.~\cite{ETMC09}-\cite{MILC10} are those taken into account in the FLAG averages of~\cite{Colangelo:2010et}. The results from~\cite{LAIHO11}-\cite{ETMC13}  are more recent than~\cite{Colangelo:2010et} and are reviewed in this talk. A superscript $*$ labels results which have been obtained in the isospin symmetric limit. The other have been extrapolated to the physical value of the up quark mass, as required for charged kaons. The corresponding isospin breaking correction is estimated to be at the $3-4\permil$ level, see Sec. 3.}
\label{tab:fkfpi}
\end{table}
Since the FLAG-1 results have been extensively discussed in that report, we will concentrate in this talk on reviewing the more recent calculations. These include two $N_f=2+1$ results, LAIHO 11~\cite{LAIHO11} (preliminary) and RBC/UKQCD 12~\cite{RBC/UKQCD12}, and three $N_f=2+1+1$ results, MILC 13~\cite{MILC13}, HPQCD 13~\cite{HPQCD13} and ETMC 13~\cite{ETMC13} (preliminary). With respect to the previous calculations, the two main improvements introduced by the recent results are the following: i) some of the new calculations (RBC/UKQCD 12, MILC 13, HPQCD 13) are based on simulations performed directly at (or very close to) the physical pion mass, thus removing in practice the uncertainty due to the chiral extrapolation in the up and down quark masses; ii) some of the new calculations (MILC 13, HPQCD 13, ETMC 13) take into account, for the first time, the dynamical effects of the charm quark in the sea.

Among the new $N_f=2+1$ calculations, the LAIHO 11 result is still preliminary, having been presented so far only at the Lattice 2011 conference~\cite{LAIHO11}. It is based on a simulation performed with domain wall fermions in the valence, a lattice regularization which exhibits good (continuum-like) chiral properties, and the staggered Asqtad quark action in the sea, with gauge field configurations produced by the MILC Collaboration. The result for $f_{K}/f_{\pi}$ has been extrapolated to the continuum limit using 3 values of the lattice spacing ($a \simeq 0.125, \ 0.09, \ 0.06$ fm) and to the physical light quark masses by simulating at values of the pion mass as low as 210 MeV in the valence and 280 MeV in the sea.

The new RBC/UKQCD 12 calculation~\cite{RBC/UKQCD12} represents a significative improvement with respect to the previous calculation of the collaboration~\cite{RBC/UKQCD10A}. They simulate domain wall fermions both in the valence and in the sea. The previous simulations were performed at only 2 values of the lattice spacing ($a \simeq 0.11 \ \rm{and} \ 0.09$ fm) with the Iwasaki gauge action. The main novelty in ~\cite{RBC/UKQCD12} is introduced by two additional ensembles generated with the novel Iwasaki+DSDR (Dislocation Suppressing Determinant Ratio) gauge action, at a coarser value of the lattice spacing ($a \simeq 0.14$~fm). The new action allows to simulate pion masses down to almost the physical value in the valence, namely $M_{\pi,min}^{val} \simeq 143$ MeV, and slightly larger masses in the sea sector, $M_{\pi,min}^{sea} \simeq 170$ MeV. With the newly generated ensembles, a much better control over both the continuum and chiral extrapolations is thus achieved.

I now discuss the lattice results for $f_{K}/f_{\pi}$ based on the new $N_f=2+1+1$ simulations. Gauge field configurations which take into account, for the first time, the effect of the charm quark in the sea have been only generated so far by the ETMC and MILC configurations. The MILC 13 calculation of $f_{K}/f_{\pi}$~\cite{MILC13} is based on simulations performed with the Highly Improved Staggered Quark  (HISQ) action, for the up, down, strange and charm quark both in the valence and in the sea. With this improved version of staggered fermions, discretization effects, which are parametrically of ${\cal O}(a^2)$, are found to be approximately 3 times smaller than those obtained with the Asqtad staggered action previously adopted by MILC in their extensive set of $N_f=2+1$ simulations. The new result for $f_{K}/f_{\pi}$ is obtained by simulating at 4 values of the lattice spacing ($a \simeq 0.15, \ 0.12, \ 0.09, \ 0.06$ fm), which allow a well controlled extrapolation to the continuum limit. Another crucial feature of this calculation is the simulation of pion masses down to the physical value. The extra species (tastes) of sea pions, which are present in simulations based on staggered fermions, have average masses in the range between 140 and 310 MeV, depending on the value of the lattice spacing. The unwanted effects of these extra tastes is cured by implementing the so-called "fourth-root procedure" and applying, at finite lattice spacing, the corrections predicted by staggered chiral perturbation theory. Finite size effects in the calculation are carefully investigated by simulating, at a fixed value of the pion mass and lattice spacing, on 3 different lattice volumes. 

On the gauge field configurations produced by MILC, with $N_f=2+1+1$ dynamical quarks and the HISQ action, is based also
the HPQCD 13 calculation~\cite{HPQCD13}, which performs however an independent analysis of the meson correlation functions. With respect to the MILC 13 calculation, HPQCD also uses heavier (than physical) up/down quark masses, 3 out of the 4 MILC lattice spacings (with the exclusion of the finest one), and a single volume for each simulated value of the pion mass. Besides the more common choice of the scale parameter $r_1$, the Wilson flow parameter $w_0$ is also used by HPQCD in order to get information on the relative values of lattice spacings
.

For this conference, I would also like to report, for the first time, the preliminary result for $f_{K}/f_{\pi}$ obtained together with my colleagues of the ETM Collaboration~\cite{ETMC13}. The ETMC simulation includes the effects of the up, down, strange and charm quark (i.e. $N_f=2+1+1$) and it is based on the Iwasaki gauge action and twisted mass fermions. We simulated at 3 values of the lattice spacing ($a \simeq 0.09, \ 0.08, \ 0.06$ fm) and several pion masses down to $M_{\pi,min} \simeq 215$ MeV. We investigated finite size effects by simulating on two lattice volumes at a fixed value of lattice spacing and pion mass.

The averages of the lattice results for $f_{K^+}/f_{\pi^+}$ collected in Table~\ref{tab:fkfpi}, evaluated separately for the calculations which include a different number of flavors in the sea, and excluding the preliminary LAIHO 11 and ETMC 13 results, lead to
\bea
\label{eq:fKfpi-3ave}
& f_{K^+}/f_{\pi^+} = 1.210\ (18) \ , & \qquad  N_f=2 \nonumber \\
& f_{K^+}/f_{\pi^+} = 1.193\ (5) \ , & \qquad  N_f=2 + 1 \nonumber \\
& f_{K^+}/f_{\pi^+} = 1.192\ (2) \ , & \qquad  N_f=2 + 1 + 1\ .
\eea
The three averages are in nice agreement within each other, showing that the sea effects of the charm quark but also of the strange quark are still smaller than other uncertainties (but the error on the $N_f=2$ average, which is based on the single ETMC 09 calculation, is significantly larger than the other two). Comparing Eq.~(\ref{eq:fKfpi-3ave}) with the FLAG averages reported in~(\ref{eq:fKfpi-FLAGave}), we see that the new RBC/UKQCD 12 calculation agrees with the previous $N_f=2+1$ results and a novelty, with respect to (\ref{eq:fKfpi-FLAGave}), is represented by new, precise determinations based on $N_f=2+1+1$ simulations at the physical point. 

A combined, weighted average of all lattice results collected in Table~\ref{tab:fkfpi} gives $f_{K^+}/f_{\pi^+} = 1.192\ (2)$, which is dominated by the single HPQCD 13 calculation. The accuracy quoted by the lattice result is at the $2\permil$ level, comparable to the one reached in the experimental determination of Eq.~(\ref{eq:exp}). The error quoted by HPQCD 13 is significantly smaller than those quoted by the other lattice calculations and, furthermore, the HPQCD 13 paper \cite{HPQCD13} has not been published yet. Therefore, before claiming a $2\permil$ accuracy for the final lattice average some additional investigation of the uncertainties is probably advisable. 

The lists of error budget reported by four of the more recent lattice calculations is presented in Table~\ref{tab:errorbud}.
\begin{table}[t]
\vspace{-0.5cm}
\centering
\includegraphics[width=0.8\textwidth]{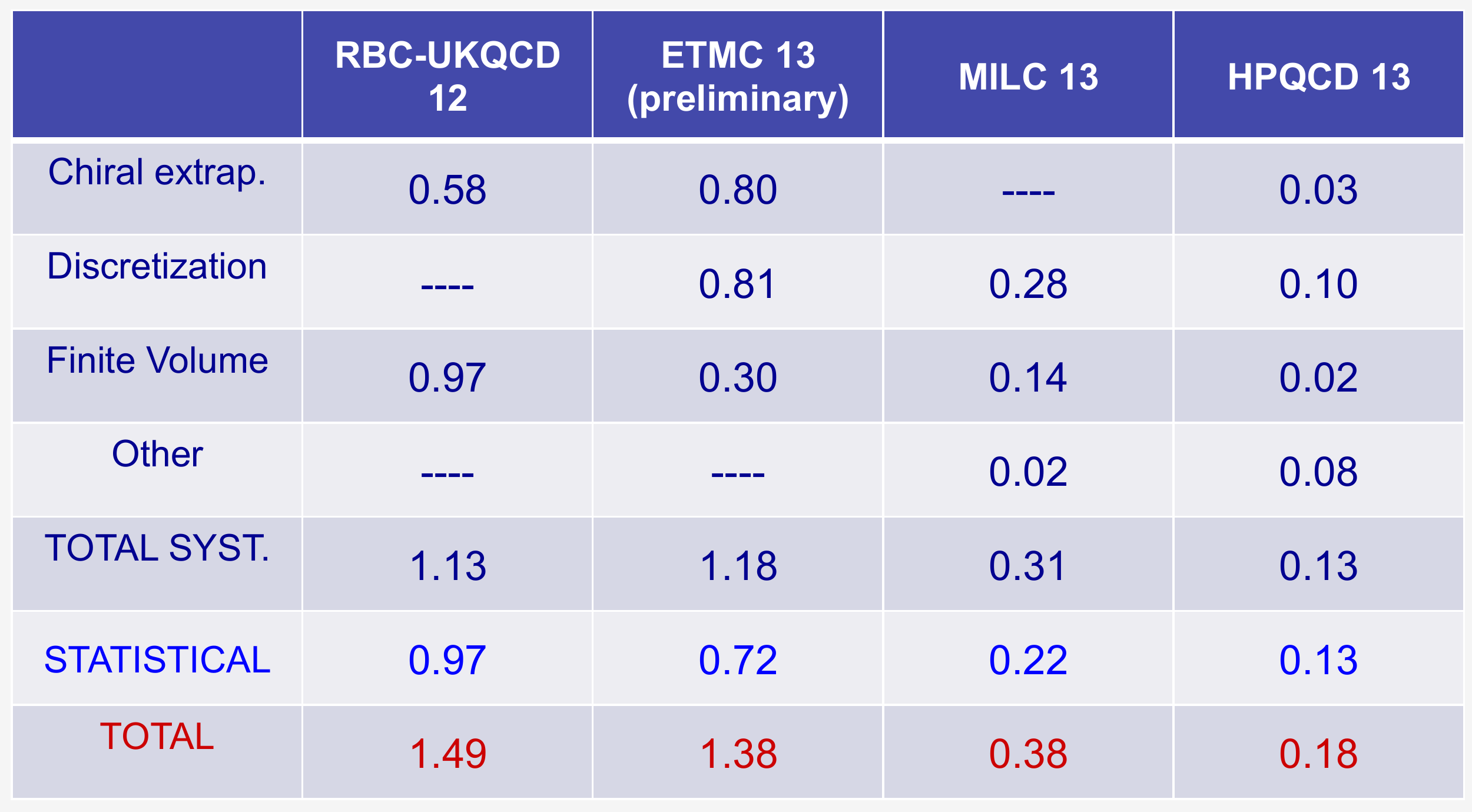}
\caption{Lists of error budget, in percent, as reported by four of the more recent lattice calculations of $f_{K}/f_{\pi}$.}
\label{tab:errorbud}
\end{table}
As can be seen from the Table, the relevant sources of systematic uncertainties in the calculations are the chiral extrapolation, discretization and finite volume effects. The associated errors turn out to be larger than the statistical ones, except for the HPQCD 13 calculation, for which they are of the same size. Adding statistical and systematic uncertainties in quadrature, the overall final errors on $f_{K}/f_{\pi}$ are estimated to be at the level of 1.5\%, 1.4\%, 0.4\% and 0.2\% for RBC/UKQCD 12, ETMC 13, MILC 13 and HPQCD 13 respectively. They differ, among the various calculations, by almost one order of magnitude.

ETMC 13 is the only one, among the four calculations, for which the lightest simulated pion mass is not at the physical value ($M_{\pi,min} \simeq 215$ MeV). Consequently, the error associated to the chiral extrapolation is large for ETMC 09, at the level of 0.8\%, corresponding to $\sim 5\%$ on $f_{K}/f_{\pi} -1$. The same error is estimated to be negligible for both MILC 13 and HPQCD 13, while it is still at the level of 0.6\% for RBC/UKQCD 12, despite the shortness of the chiral extrapolations required in that calculation. In the RBC/UKQCD 12 calculation, which eventually quotes an overall final uncertainty similar to the one quoted by ETMC 13, discretization errors are evaluated to be negligible, and the systematic uncertainty is dominated instead by finite volume effects. 

Of particular interest in Table~\ref{tab:errorbud} is the comparison of error budgets between MILC 13 and HPQCD 13. First of all, these are the two lattice results for $f_{K}/f_{\pi}$ which quote the smallest uncertainties, at the permille level, with the HPQCD 13 error being two times smaller than the MILC 13 one. In addition, both calculations use the ensembles of gauge field configurations generated by MILC, and simulate with the same lattice action for valence quarks as well. As the lattice setup and also the number of gauge configurations are similar in the two calculations, one would have expected uncertainties of similar size whereas, as indicated in Table~\ref{tab:errorbud}, they differ by about a factor 2. By looking in more details at the papers of the two collaborations, personally I have not been able to better understand the origin of this difference. The MILC collaboration performs a dedicated study of finite volume effects, based on 3 lattice sizes at fixed pion mass and lattice spacing, while HPQCD uses only a single volume for each ensemble. In spite of this, the finite size error quoted by HPQCD is approximately 7 times smaller than the one quoted by MILC. The analysis of discretization errors is similarly puzzling. The HPQCD calculation uses only 3 out of the 4 values of the lattice spacing used by MILC, by not simulating at the finest one, i.e. the one closest to the continuum limit, but the uncertainty quoted for discretization effects by HPQCD is almost 3 times smaller than the one quoted by MILC. Also the statistical error quoted by HPQCD is almost a factor 2 smaller than the one obtained by MILC, despite the statistical samples of gauge configurations being similar in size in the two calculations. Not having been directly involved in the numerical analyses, I cannot really judge whether the different precision quoted by the two calculation is due to an underestimate of the uncertainties by HPQCD or an overestimate by MILC, or perhaps both. The accuracy claimed by the two calculation is very high in both cases (with respect to other lattice results). Since the lattice prediction for $f_{K}/f_{\pi}$ plays a crucial role in the determination of the CKM matrix element $V_{us}$, which, in turn, has a relevant impact on many phenomenological analyses of particle physics, I prefer to stay on the more conservative side, relying on the uncertainty quoted by MILC. Thus, as a final lattice estimate for $f_{K^+}/f_{\pi^+}$, I quote
\be
\label{eq:fKfpi-ave}
f_{K^+}/f_{\pi^+}= 1.193\ (4) \ . 
\ee
The accuracy is at the level of $4\permil$, corresponding to a $2\%$ precision on $f_{K^+}/f_{\pi^+}-1$. This error is approximately 2 times larger than the experimental uncertainty given in Eq.~(\ref{eq:exp}).

Of high phenomenological interest is also the lattice determination of the kaon decay constant $f_{K}$, besides the ratio $f_{K}/f_{\pi}$, since its absolute value is relevant for example in the analysis of $K-\bar K$ mixing, together with the one of the bag parameter $B_K$. The recent lattice results for $f_{K}$ are also collected in Table~\ref{tab:fkfpi}. Note that, since the experimental value of the pion decay constant $f_{\pi}$ is used as input in some of the lattice calculations in order to fix the lattice scale, the two sets of results for $f_{K}$ and $f_{K}/f_{\pi}$ in the Table are not always independent. From these results, as a final lattice average for $f_{K^+}$ I quote
\be
\label{eq:fK-ave}
f_{K^+} = 155.7\ (0.6) \ {\rm MeV} \ ,
\ee
which exhibits the same accuracy, at the $4\permil$ level, of the result in Eq.~(\ref{eq:fKfpi-ave}) .

\section{Isospin breaking effects}
The experimental analysis of leptonic kaon decays, which leads to the measurement quoted in Eq.~(\ref{eq:exp}), refers to the decays of charged kaons (and pions). On the other hand lattice results, until very recently, have been obtained in the isospin symmetric limit, i.e. by neglecting the mass difference between the up and down quarks and the electromagnetic interactions. Parametrically, the size of isospin breaking corrections is controlled by
\be
\label{eq:IBeff}
{\cal O}\left(\frac{m_d-m_u}{\Lambda_{QCD}} \right) \sim 1\% \quad \ {\rm and} \quad 
{\cal O}\left( \alpha_{em} \right) \sim 1\% \ .
\ee
Thus, given the level of precision reached by both experimental and theoretical determinations in flavor physics, these effects cannot be neglected. In this section, we briefly discuss the recent progress achieved in evaluating the isospin breaking effects using lattice QCD by concentrating on the case of leptonic kaon decays. For a more general discussion on lattice studies of isospin breaking effects, see the dedicated talk by A.~Portelli at this conference~\cite{Portelli}. 

A net separation between QCD ($\sim m_d-m_u$) and electromagnetic ($\sim \alpha_{em}$) isospin breaking effects, as suggested by Eq.~(\ref{eq:IBeff}), is purely conventional, since the masses of the up and down quarks also receive (ultraviolet divergent) electromagnetic contributions. Nevertheless, once a given definition for the quark masses in pure QCD has been chosen, the two effects can be identified. The ratio of the leptonic decay rates, for instance, can be expressed in the form
\be
\label{eq:rates}
\frac{\Gamma(K^{+} \to \ell \nu_\ell (\gamma))}{\Gamma(\pi^{+} \to \ell \nu_\ell (\gamma))} = 
\frac{|V_{us}|^2}{|V_{ud}|^2} \left( \frac{ f_{K^+} } {f_{\pi^+} } \right) ^2 
\frac{m_{K^+} (1- m_\ell^2/m_{K^+}^2)}{m_{\pi^+} (1- m_\ell^2/m_{\pi^+}^2)} \, \left( 1 + \delta_{EM} \right) \ ,
\ee
where the decay constants $f_{K+}$ and $f_{\pi+}$ are defined for charged kaon ($u \bar s$) and pion ($u \bar d$) in pure QCD, for defined values of the quark masses. The factor $( 1 + \delta_{EM})$ in Eq.~(\ref{eq:rates}) then accounts for the electromagnetic corrections to the decay rates. Note, in particular, that the meson decay constants can be {\em only} defined within pure QCD, since in the presence of electromagnetic interactions the relevant matrix elements of the weak axial charged currents are not even QED gauge invariant.

Within QCD, the isospin breaking effects in the ratio $f_{K^+}/f_{\pi^+}$ can be also expressed as
\be
\frac{ f_{K^+} } {f_{\pi^+}} = \frac{ f_{K} } {f_{\pi} }  \, \left( 1 + \delta_{SU(2)} \right) \ , 
\ee
where $f_{K}$ and $f_{\pi}$ represent the decay constants in the isospin symmetric limit, i.e. evaluated for up and down quarks with a common mass $m_{ud}=(m_d+m_u)/2$. The isospin breaking correction $\delta_{SU(2)}$ can be then evaluated as a series expansion in powers of the mass difference $\Delta m_{ud}=(m_d-m_u)/2$. Usually, retaining only the first term in this expansion is sufficient, since as previously noted the leading correction is already of the order of 1\%.

The interchange symmetry between up and down quarks in the $\pi$ mesons guarantees that the decay constant $f_{\pi^+}$ is an even function of $\Delta m_{ud}$, so that
\be
f_{\pi^+} = f_\pi \left( 1 + {\cal O}(\Delta m_{ud}^2) \right) \ .
\ee
Therefore, at the leading order, the SU(2) breaking correction $\delta_{SU(2)}$ is determined only by the kaon decay constants and one has
\be
\label{eq:IBexp}
f_{K^+} = f_K \left( 1 + \delta_{SU(2)}) \right) = f_K + \left(\frac{\partial f_{K^+}}{\partial \Delta m_{ud}}\right)_0 \Delta m_{ud} + \ldots \ .
\ee

A strategy to accurately evaluate isospin breaking corrections on the lattice has been recently proposed~\cite{IB1,IB2}. It is based on the expansion of the Euclidean functional integral in powers of the of the up-down quark mass difference $\Delta m_{ud}$~\cite{IB1} and, when electromagnetic corrections have to be taken into account, in powers of the coupling $\alpha_{em}$~\cite{IB2}. The main observation is that the smallness of the isospin breaking correction is determined by the tiny values of the up-down quark mass difference and of the electromagnetic coupling, but the coefficients of the expansion instead, as the slope $(\partial f_{K^+}/\partial \Delta m_{ud})_0$ in Eq.~(\ref{eq:IBexp}), have the usual size determined by the strong interaction scale, and can be directly evaluated on the lattice with the required accuracy.

\begin{figure}
\vspace{-0.5cm}
\centering
\includegraphics[width=0.6\textwidth]{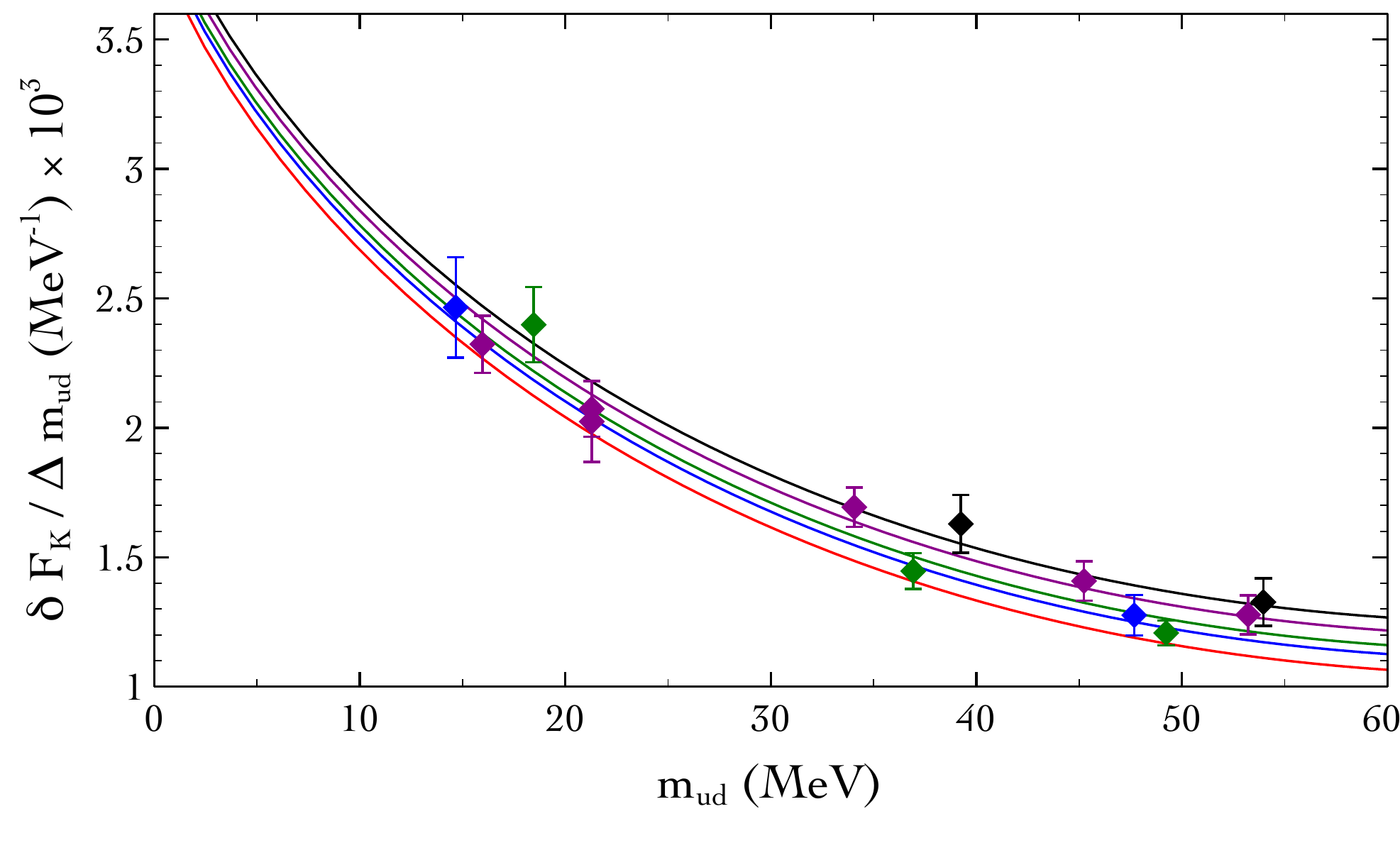}
\vspace{-0.5cm}
\caption{\it Lattice results for the isospin breaking slope $(\partial f_{K^+}/\partial \Delta m_{ud})_0$ as a function of the average up-down quark mass $m_{ud}$, from Ref.~\cite{IB1}. Black, magenta, green and blue points correspond to different values of lattice spacing, and the red line is the result of the continuum extrapolation. The physical value of the slope is given by the extrapolation to the physical value of the light quark mass, $m_{ud}\simeq 3.5$ MeV.}
\label{fig:dfK}
\end{figure}

The result obtained for $(\partial f_{K^+}/\partial \Delta m_{ud})_0$, as a function of the average up-down quark mass, is shown in Fig.~\ref{fig:dfK} taken from Ref.~\cite{IB1}, which is the first lattice calculation of the isospin breaking correction to the kaon decay constant.
Once combined with a lattice determination of the up-down mass difference, which required also an evaluation of the electromagnetic corrections to the charged and neutral kaon masses, it leads to the result~\cite{IB2}
\be
\label{eq:deltasu2}
\delta_{SU(2)} = \frac{1}{f_K} \left(\frac{\partial f_{K^+}}{\partial \Delta m_{ud}}\right)_0 \Delta m_{ud} = -0.40 (3) (2) \% \ .
\ee
The lattice result is somewhat smaller than the typical 1\% size of isospin breaking corrections but it is larger than the estimate obtained in Ref.~\cite{dfK-ChPT} using chiral perturbation theory, namely $\delta_{SU(2)}$= -0.21 (6) \%. A lattice estimate of the $\delta_{SU(2)}$ has been also quoted by the HPQCD collaboration~\cite{HPQCD13}, by studying the dependence of $f_K$ on the values of the valence light quark mass. This dependence is directly related to $\delta_{SU(2)}$, since the isospin breaking corrections due to the sea up and down quark masses are quadratic in $\Delta m_{ud}$, and the $K$ mesons  contain only one up or down quark in the valence. Thus, in this specific case, the derivative $(\partial f_{K^+}/\partial \Delta m_{ud})_0$ coincides with the derivative with respect to the single valence light quark mass, which can be estimated on the lattice by simulating at different values of this mass. Proceeding in this way, some of the lattice results for the kaon decay constants which are collected in Table~\ref{tab:fkfpi} have already taken into account the leading isospin breaking correction, by extrapolating the numerical data in the light valence quark mass down to the $up$ quark mass, as required for charged kaons. From this extrapolation, HPQCD 13 has derived a numerical estimate of the slope~\cite{HPQCD13}, which corresponds to $\delta_{SU(2)}$= -0.27 (6) \%. This result is in between the one obtained from the direct lattice calculation of the slope~\cite{IB1}, see Eq.~(\ref{eq:deltasu2}), and the chiral perturbation theory prediction.

The studies of isospin breaking effects are becoming at present an important task for lattice QCD calculations, and results with increasing accuracy are expected to be produced in the near future. It is also worth noting that while the relative errors on isospin breaking effects are still large at present, at the level of 20-30\%, they actually represent uncertainties on tiny corrections, resulting in an accuracy on the corresponding hadronic parameters which are typically at the permille level. A much more challenging task for the lattice calculations is represented by the evaluation of the electromagnetic corrections to the decay rates, represented by $\delta_{EM}$ in Eq.~(\ref{eq:rates}). A recent estimate of this correction is $\delta_{EM}$= -0.69 (17) \% ~\cite{dfK-ChPT}, obtained by using chiral perturbation theory supplemented by large-$N_c$ models. Needless to say, a first principle, lattice determination of this correction would be clearly very welcome.

\section{Lattice calculations of $\mathbf{f_{+}(0)}$}
The last topic of this talk is the lattice determination of the kaon semileptonic vector form factor $f_+(0)$ which, once combined with the experimental measurement of the product $|V_{us}|f_+(0)$ given in Eq.~(\ref{eq:exp}), allows for a precise, direct determination of the CKM matrix element $V_{us}$.

A strategy to evaluate on the lattice the kaon semileptonic vector form factor with the percent accuracy required for phenomenology has been developed in Ref.~\cite{Becirevic:2004ya}, which obtained the result $f_+(0)=0.960\ (9)$, in the quenched approximation ($N_f=0$). This result has been later confirmed by unquenched lattice calculations, and the best lattice estimate of the form factor reported recently by FLAG-1~\cite{Colangelo:2010et} is:
\be
\label{eq:fp0-FLAGave}
f_+(0)=0.956\ (8)  \ , \qquad  \textrm{FLAG-1} \ \  - \ \ N_f= 2 \ \ \& \ \ 2 + 1 \ .
\ee
In this talk I will review the new calculations, appeared after FLAG-1, and update the lattice average. It is also worth noting that the lattice results for $f_+(0)$ tend to be smaller than the predictions of analytical models (for a more detailed comparison see e.g. Ref.~\cite{Colangelo:2010et}). 

\begin{figure}
\vspace{-0.8cm}
\centering
\includegraphics[width=0.8\textwidth]{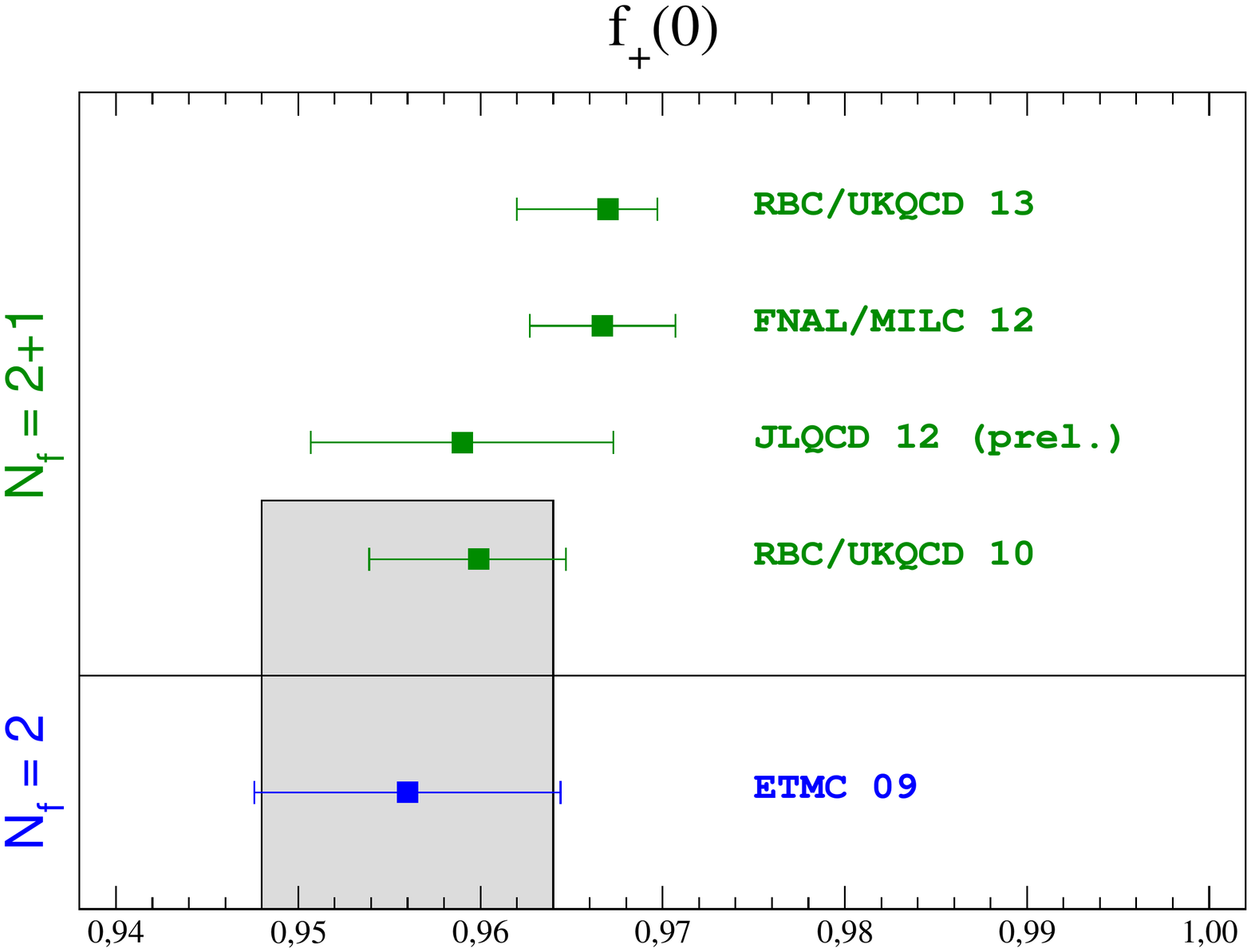}
\vspace{-1.0cm}
\caption{\it Lattice results for the kaon semileptonic form factor $f_{+}(0)$. The results shown in the lower side of the plot are those taken into account for the best estimate in FLAG-1~\cite{Colangelo:2010et}. This estimate is shown by the gray band. The new results, appeared after the FLAG-1 compilation, are shown on the upper side of the plot.}
\label{fig:fp0}
\end{figure}

The FLAG-1 estimate was based on two lattice calculations only, ETMC 09A~\cite{ETMC09A} and RBC/ UKQCD 10~\cite{RBC/UKQCD10}. In the $N_f=2$ calculation of ETMC 09A all sources of systematic uncertainties were carefully kept under control, the main uncertainty of that calculation being represented by the quenching of the strange quark. The RBC/UKQCD 10 calculation  was based on a $N_f=2+1$ simulation but performed with rather large values of pion masses ($M_{\pi,min} \simeq 330$ MeV) and at a single, rather coarse value of the lattice spacing ($a \simeq 0.11$ fm), which prevented the extrapolation of the lattice result to the continuum limit. Both the ETMC 09A and RBC/UKQCD 10 results are shown in the lower part of the plot in Fig.~\ref{fig:fp0}, together with the FLAG-1 best estimate~(\ref{eq:fp0-FLAGave}) represented by the gray band.

After the FLAG-1 compilation, three new determinations of $f_+(0)$ have been produced: JLQCD 12~\cite{JLQCD12} (preliminary), FNAL/MILC 12~\cite{FNAL/MILC12} and RBC/UKQCD 13~\cite{RBC/UKQCD13}~\footnote{Only a preliminary result~\cite{RBC/UKQCD12p} of RBC/UKQCD 13 was available at the time of the conference. A discussion of Ref.~\cite{RBC/UKQCD13}, which has recently appeared, is included in the written version of this talk for completeness.}. These results are shown in the upper side of the plot in Fig.~\ref{fig:fp0} and the whole list of recent results is collected in Table~\ref{tab:fp0}.
\begin{table}[t]
\vspace{-0.5cm}
{\footnotesize
\begin{tabular*}{\textwidth}{l @{\extracolsep{\fill}} c c c}
\hline\hline \\[-0.2ex]
  & {Ref.} & {$N_f$} & {$f_{+}(0)$}
\\[0.2ex] \hline \hline \\[-0.2ex]
RBC/UKQCD 13 & \cite{RBC/UKQCD13} & 2+1 & 0.9670 ($^{+27}_{-50}$)
\\[2.0ex] \hline \\[-2.0ex]
FNAL/MILC 12 & \cite{FNAL/MILC12} & 2+1 & 0.9667 (40)
\\[2.0ex] \hline \\[-2.0ex]
JLQCD 12 (prel.) & \cite{JLQCD12} & 2+1 & 0.9590 (83)
\\[2.0ex] \hline \\[-2.0ex]
RBC/UKQCD 10 & \cite{RBC/UKQCD10} & 2+1 & 0.9599 ($^{+48}_{-60}$)
\\[2.0ex] \hline \hline \\[-2.0ex]
ETMC 09A & \cite{ETMC09A} & 2 & 0.9560 (84)
\\[2.0ex] \hline \hline
\end{tabular*}
}
\caption{\it Lattice results for the kaon semileptonic vector form factor $f_{+}(0)$. The results from Ref.~\cite{ETMC09A} and \cite{RBC/UKQCD10} appeared before FLAG-1~\cite{Colangelo:2010et} and are those taken into account by FLAG-1 in order to derive a best estimate of the form factor. The results from Refs.~\cite{JLQCD12}-\cite{RBC/UKQCD13}  are new.}
\label{tab:fp0}
\end{table}

The JLQCD 12 determination~\cite{JLQCD12} of the vector form factor has been presented at the Lattice 2012 conference and it is still preliminary. It has been obtained by simulating with overlap fermions, which is the lattice discretization of the QCD quark action which exhibits the best (continuum like) chiral properties. However, the result has been obtained so far by using only relatively large pion masses ($M_{\pi,min} \simeq 290$ MeV) and at a single value of the lattice spacing ($a \simeq 0.11$ fm).

The FNAL/MILC 12~\cite{FNAL/MILC12} represents the first $N_f=2+1$ lattice calculation of $f_+(0)$ in which all source of systematic uncertainties have been kept under control in a satisfactory way. It is based on a simulation with the staggered HISQ action for the valence quarks and the {\it old} Asqtad action for the sea. It uses only 2 values of the lattice spacing ($a \simeq 0.12$ and 0.09 fm) and pion masses with $M_{\pi,min} \simeq 260$ MeV. The vector form factor $f_+(0)$ is evaluated by computing the scalar form factor $f_0(q^2)$ from the matrix element of the scalar density on the lattice (in this way a determination of the vector current renormalization constant is not required), and using the identity $f_+(0)=f_0(0)$ which is valid at $q^2=0$. It is also interesting to report that the FNAL/MILC Collaboration is currently working on a much improved lattice calculation of the semileptonic kaon form factor~\cite{FNAL/MILC12p}, which accounts for the sea effects of the charm quark ($N_f=2+1+1$), uses the HISQ action for both valence and sea, and simulate pion masses directly at the physical point.

The second new calculation of $f_+(0)$, appeared after FLAG-1, is RBC/UKQCD 13~\cite{RBC/UKQCD13}, which is also based on a $N_f=2+1$ simulation. With respect to the previous calculation of the collaboration, RBC/UKQCD 10~\cite{RBC/UKQCD10}, it improves on two important aspects. It is based on ensembles of gauge configurations generated at 3 (rather than one) values of the lattice spacing ($a \simeq 0.14, \ 0.11, \ 0.09$ fm), and simulates pion masses as low as $M_{\pi,min} \simeq 170$ MeV (rather than 330 MeV). As already  discussed in Sec.~1, the latter achievement has been made possible by the use of the new Iwasaki+DSDR gauge action in the simulation at the coarser value of the lattice spacing (the lattice setup adopted for this calculation is similar to the one used in the RBC/UKQCD 12 determination of $f_K/f_\pi$). The final result for $f_+(0)$ quoted by RBC/UKQCD 13 is accompanied by a largely asymmetric error, see Table~\ref{tab:fp0}. It follows from the choice of the authors to select as a preferred fit for the chiral extrapolation of the form factor the one based on a polynomial expansion,  with respect to the functional form based on chiral perturbation theory\footnote{In order to evaluate the average of lattice results for $f_+(0)$ given in Eq.~(\ref{eq:fp0-ave}), I will symmetrize the error quoted by RBC/UKQCD 13.}.

As can be also seen from Fig.~\ref{fig:fp0}, the new results produced by FNAL/MILC 12 and RBC/UKQCD 13 quote central values of $f_+(0)$ which are slightly larger than the one given by the FLAG-1 best estimate. In order to derive an updated average for $f_+(0)$, I exclude from the results listed in Table~\ref{tab:fp0} the RBC/UKQCD 10 determination, which has been superseded by RBC/UKQCD 13, and the JLQCD 12 result, which is still preliminary. The $N_f=2$ determination of ETMC 09 is averaged together with the other $N_f=2+1$ results since in the former the error due to the quenching of the strange quark, which for $f_+(0)$ only enters at the NNLO in chiral perturbation theory, has been taken into account, together with the other uncertainties. In this way, I obtain the updated average
\be
\label{eq:fp0-ave}
f_+(0)=0.965\ (3)  \ .
\ee
The accuracy is at the level of $3\permil$, corresponding to a $9\%$ precision on $1-f_+(0)$. The error is only slightly larger than the experimental one given in Eq.~(\ref{eq:exp}).

\section{Conclusions}
In this talk I have reviewed the recent lattice QCD calculations for the ratio of decay constants $f_{K^+}/f_{\pi^+}$ and the semileptonic kaon vector form factor $f_+(0)$, which are the two hadronic parameters playing a crucial role in the determination of the CKM matrix element $V_{us}$.

The lattice results for $f_{K^+}/f_{\pi^+}$ have reached today a sub-percent accuracy. State of the art simulations include the effect the up, down, strange and charm quarks in the sea ($N_f=2+1+1$ simulations) and the uncertainty due to the chiral extrapolation has been eliminated in several calculations, with simulations performed directly at (or very close to) the physical pion mass. In this talk, I have also critically examined in some detail the error budget lists quoted by the more recent calculations. The present level of accuracy reached by the lattice calculations of the decay constants renders isospin breaking corrections definitively important. They have started being computed on the lattice for the QCD part, whereas the lattice evaluation of long distance electromagnetic corrections to the leptonic decay rates is presumably feasible, but definitively more challenging.

As far as the semileptonic kaon form factor $f_+(0)$ is concerned, the list of lattice calculations with fully controlled systematic uncertainties is less numerous than for the decay constants, but new, accurate results have recently appeared and further, improved calculations are in progress, by several lattice collaborations. 

Updated lattice averages for $f_{K^+}/f_{\pi^+}$ and $f_+(0)$ are presented in Eqs.~(\ref{eq:fKfpi-ave}) and (\ref{eq:fp0-ave}) respectively.

\section*{Acknowledgments}
It is a pleasure to warmly thank the organizers of KAON13, and in particular Monica Tecchio, for the nice and stimulating atmosphere of the conference. I also thank Silvano Simula and Cecilia Tarantino for a careful reading of this draft.

\end{document}